\documentclass[a4paper]{article} 

\usepackage{graphicx}
\usepackage{dcolumn}
\usepackage{bm}
\usepackage{amsmath}
\usepackage{amssymb}

\begin{document}

\title{Data assimilation as a nonlinear dynamical systems problem: Stability and convergence of the prediction-assimilation system}

\author{Alberto Carrassi$^1$, Michael Ghil$^{2,3}$, Anna Trevisan$^4$, and Francesco Uboldi$^5$} 
\maketitle

[1]: Institut Royal M\'et\'eorologique de Belgique, Bruxelles, Belgique$^*$

[2]: \'Ecole Normale Sup\'erieure, Paris, France

[3]: University of California, Los Angeles, CA 90095-1567, USA

[4]: ISAC-CNR, Bologna, Italy

[5]: Novate Milanese, Italy\\

{\small $^*$Corresponding author e-mail: a.carrassi@oma.be}


\begin{abstract}
We study prediction-assimilation systems, which have become routine 
in meteorology and oceanography and are rapidly spreading to other 
areas of the geosciences and of continuum physics. The long-term, 
nonlinear stability of such a system leads to the uniqueness of its 
sequentially estimated solutions and is required for the convergence 
of these solutions to the system's true, chaotic evolution. The key 
ideas of our approach are illustrated for a linearized Lorenz system. 
Stability of two nonlinear prediction-assimilation systems from 
dynamic meteorology is studied next via the complete spectrum of 
their Lyapunov exponents; these two systems are governed by a large 
set of ordinary and of partial differential equations, respectively. 
The degree of data-induced stabilization is crucial for the 
performance of such a system. This degree, in turn, depends on two 
key ingredients: (i) the observational network, either fixed or 
data-adaptive; and (ii) the assimilation method.

\end{abstract}



\noindent {\bf Physical systems --- in nature, the laboratory or industry --- can 
only be measured at a limited number of points in space and time. 
Estimating the state of a nonlinear dynamical system from 
partial and noisy observations is therefore crucial in applied 
physics and engineering \cite{Jaz70,gelb74}. In numerical weather
and ocean prediction, this classical estimation problem goes under 
the name of {\it data assimilation} \cite{bengtsson81, GhilMal91}; as
data assimilation is spreading rapidly to other fields of the 
geosciences and of continuum physics, it is important to better grasp 
its fundamental theoretical aspects. In practice, a data assimilation 
algorithm has to provide the best-possible estimate of the evolving 
state of the system, using the observations available and the 
equations governing the system's time evolution \cite{talagrand97}. 
In this paper, we examine the long-term stability of 
the set of modified equations that are referred to as the {\it 
prediction-assimilation system}, in the case in which the original 
physical system is fully nonlinear and chaotic. }

\section{Introduction and motivation}

\subsection{Background}
The complete solution of the {\it filtering and prediction problem} in
sequential-estimation theory \cite{Jaz70,gelb74} is given by the 
probability density function (PDF) of the unknown state, conditioned 
on the observations.
Given the correct initial PDF and assuming that the system noise and 
observational noise are white, normally distributed, mutually 
uncorrelated and known, the PDF's time evolution can be predicted by 
the Fokker-Planck equation \cite{Jaz70}. In the case of a continuous 
stochastic dynamical system, with partial observations distributed at 
discrete times, an ideal data assimilation scheme would solve the 
Fokker-Planck equation for the time interval between observations and 
modify the PDF by using all observations when available.

  The fundamental difficulty of this approach relates to the high 
dimension of the state space, which
  makes it impossible in practice to obtain the initial PDF, let alone 
compute its time evolution.
In the case of linear dynamics and of observations that are linearly 
related to the system's state variables, the PDF is fully 
characterized by its first and second moments, {\it i.e.} by the mean 
and covariance respectively.
The optimal solution of the data assimilation problem in this linear 
setting is provided by the Kalman filter (KF) equations that describe 
the time evolution of both the mean and the covariance 
\cite{Jaz70,gelb74,bengtsson81,GhilMal91,talagrand97,kalman60,kalmanbucy61}.
The time-dependent error-covariance matrix depends, in this linear case,
only on the observational error statistics that are part of the 
problem statement, and not on the actual observations  \cite{Jaz70}.

Thus, in the case of linear dynamics, a linear observation operator, 
and observational and system noise that are both Gaussian, white in 
time and mutually uncorrelated, the KF equations give the optimal 
linear estimate of the state of the system by propagating the 
associated error covariances, along with the state estimates.
In the nonlinear case, the situation is vastly more difficult and the 
PDF cannot be described by a finite set of parameters. A 
straightforward way of extending the linear results to the nonlinear 
case is given by the Extended Kalman filter (EKF) 
\cite{GhilMal91,talagrand97,Ghil97,kalnay03}.

In the EKF the tangent linear operator is used for predicting the 
approximate error statistics, while the state evolves according to 
the full, nonlinear equations. The computational cost of the EKF, 
though, is still prohibitive in many realistic circumstances. To 
alleviate this problem, a number of authors have studied reduced-rank 
approximations of the full EKF 
\cite{Ghil97,TodlingCohn94,Tippetetal2000,Fukumori02}
that allow a reduction of its computational cost, while maintaining a 
satisfactory accuracy of the sequential estimates. A Monte Carlo 
approach, referred to as Ensemble Kalman filter, has also proven 
effective in reducing the computational cost associated with the full 
EKF \cite{kalnay03,evensen94,keppenne-rienecker02}.

Early theoretical work on the observability and stability of 
distributed-parameter systems ({\it i.e.}, systems governed by 
coupled partial differential equations) was confined, by-and-large, 
to linear dynamics and to predetermined observations. In the case of 
linear, lumped-parameter systems ({\it i.e.}, systems governed by 
coupled ordinary differential equations) and observations that are 
both linear and discrete in time, a sufficient condition for the KF 
solution to be stable is given by the observability of its dynamics 
\cite{Jaz70,kalmanbucy61}. Cohn and Dee \cite{CohnDee88} have shown, 
in the linear, infinite-dimensional case of distributed-parameter 
systems that it is important to consider observability in the context 
of the discretized system, and that this observability implies 
stability of the data-assimilation problem.

The concepts of observability and stability for nonlinear chaotic 
systems are closely related to other areas of dynamical system 
theory, namely {\it controlling chaos} and {\it synchronization}. In 
the control of chaos, a significant modification of the system's 
behavior is achieved by small variations in time of some parameter. 
Originally devised to stabilize unstable periodic orbits 
\cite{Ott_etal_1990}, this approach has been generalized to force a 
given dynamical system to achieve other desiderable types of 
behavior, wheter stationary, periodic or chaotic 
\cite{Boccaletti_2000}. In the present context, synchronization of 
chaotic systems means essentially using an adaptive coupling to have 
a "slaved" system track the motion of a driver or "master" system 
\cite{Boccaletti_2000,ChenKurths07}. 

At the core of both chaos control and syncronization lies the 
stability problem. In the former, the time-dependent control  has to 
be chosen so as to stabilize the motion. In the latter, the stability 
of the synchronized motion is a necessary condition for achieving 
such a motion.

Interesting applications of both chaos control and synchronization to 
geophysical problems include the work of Tziperman et al. \cite{Tziperman_etal_1997} 
on stabilizing an unstable periodic orbit in a fairly realistic El 
Ni\~no model governed by a set of coupled, nonlinear partial 
differential equations, as well as that of Duane et al. \cite{Duane_Tribbia_2002} 
on meteorological teleconnections between the Atlantic and Pacific 
sectors of the Northern Hemisphere. Moreover, the relation between 
synchronization and data assimilation has been investigated by Duane 
et al. \cite{Duane_etal_2006} and by Yang et al. 
\cite{Yang_etal_2006}.

\subsection{The present approach}

In a chaotic system, initial errors grow within the system's unstable 
subspace. Trevisan and Uboldi \cite{TU,UT} considered fully nonlinear 
and possibly chaotic dynamics and proposed, in the context of meteorological 
data assimilation, to detect and eliminate the unstable components of 
the forecast error. They showed that those observations that help 
detect such instabilities maximize error reduction in the state 
estimates. Ghil \cite{Ghil97} and associates (see references there) 
had already shown that, in meteorological and oceanographic data 
assimilation, the number of observations necessary to track an 
unstable flow is comparable to the number of the flow's dominant 
degrees of freedom, while Carrassi et al. \cite{CTU} showed that one 
can improve on this estimate, since the requisite number of ``tracking 
observations" is closely related to the number and magnitude of the 
system's positive Lyapunov exponents.

In this paper, we examine the long-term stability of 
prediction-assimilation systems. This stability is essential for the 
performance of data assimilation methods and the convergence of their 
sequential estimates to the correct evolution of the underlying 
physical system. We develop a theoretical framework for the study of 
this long-term stability, and present a theorem that, under certain 
simplifying assumptions, provides rigorous conditions for the 
stability of the prediction-assimilation system. Within this 
framework, we present an approach that optimizes the convergence of 
the estimates to the correct solution, and apply it to two 
meteorological models of increasing complexity.

The paper is organized as follows. In Sect. 2 we describe the 
formulation of data assimilation for nonlinear, chaotic dynamics, 
with particular emphasis on the proposed approach of Assimilation in 
the Unstable Subspace (AUS). Section 3 presents first  the theorem, 
its proof and an illustrative numerical example; this illustration is 
followed by numerical results on the two nonlinear models, one 
governed by a large system of ordinary differential equations, the 
other by a system of coupled partial differential equations. 
Concluding remarks appear in Sect. 4.

\section{Data assimilation for chaotic dynamics}
We concentrate here on dynamical systems that are perfectly 
deterministic but chaotic; the role of explicit stochastic forcing, 
which may represent unresolved scales of motion, will be considered 
in future work. Without loss of generality, we write the system as a 
mapping from an arbitrary initial state at time $t_{0}$ to a later 
time $t$:
\begin{equation}
{\bf x}(t) = {\cal M} ({\bf x}(t_0)),
\end{equation}
where ${\bf x}$ is the $n$-dimensional state vector and ${\cal M}$ 
the nonlinear evolution operator. Given the initial state ${\bf 
x}(t_0)={\bf x}_0$, Eq. (1) will predict the state at future times 
$t$. However, due to the chaotic nature of the system, initial errors 
will amplify in time, thus setting a limit to the system's 
predictability.

The tangent linear equations describing the evolution of 
infinitesimal perturbations $\delta{\bf x}$ relative to an orbit of 
Eq. (1) can be written as:
\begin{equation}
\delta{\bf x}(t) = {\bf M} \delta{\bf x}(t_0),
\end{equation}
where ${\bf M}={\bf M}({\bf x}(t_0),t-t_0)$ is the linearized 
evolution operator associated with ${\cal M}$, along the portion of 
trajectory between $t_0$ and $t$.
A chaotic system possesses one or more positive Lyapunov exponents, 
while their full spectrum characterizes the system's  stability 
properties; these properties are crucial for the filtering, as well 
as for the prediction problem \cite{Jaz70, Ghil-etal81}.

Suppose we seek an estimate of the state of this chaotic dynamical 
system from a set of noisy observations, given at discrete times 
$t_k\geq t_0$, $k\in\{1,2,...\}$,
\begin{equation}
{\bf y}^0_k = {\cal H} ({\bf x}_k) + \varepsilon^o_k;
\end{equation}
here ${\bf y}^0_k$ denotes the $p$-dimensional observation vector, 
${\bf x}_k$ the unknown true state, and $\varepsilon^o_k$ is the 
observational error, all at time $t_k$, while ${\cal H}$ is the (possibly 
nonlinear) observation operator.
The observational error is assumed to be Gaussian with zero mean and 
known covariance matrix ${\bf R}$. We consider the underdetermined 
situation $p \le n$; typically $p << n$ in applications.

To obtain an estimate of the state of the system, referred to in 
meteorological practice as the ${\it analysis}$ ${\bf x}^a$, one 
combines all available observations at $t_k$ with the {\it 
background} information, which consists of the forecast state at 
$t_k$. This {\it update} is given by the analysis equation:
\begin{equation}
{\bf x}^a_k = ({\bf I} - {\bf K}_k{\cal H}) {\bf x}^f_k+{\bf K}_k{\bf y}^o_k,
\end{equation}
where ${\bf x}^f_k$ indicates the {\it forecast} state, and ${\bf 
K}_k$ is the {\it gain matrix} at time $t_k$.
We use here the unified notation \cite{ide97} for meteorological and 
oceanographic data assimilation. In most sequential algorithms, the 
analysis equation has the form (4); such algorithms include the 
extended Kalman filter (EKF), as well as so-called {\it optimal 
interpolation} and other practical data assimilation schemes 
\cite{bengtsson81, GhilMal91,kalnay03}. Computing the optimally 
feasible ${\bf K}$ is at the heart of the {\it sequential-estimation 
approach} to filtering and prediction
\cite{Jaz70, gelb74, kalman60}.

The analysis state at time $t_k$ is obtained by applying the update 
(4) at this time to the forecast state ${\bf x}^{f}_k$ given by the 
nonlinear model evolution (1) of the analysis at the previous 
observation time $t_{k-1}$:
\begin{equation}
{\bf x}^a_k = ({\bf I} - {\bf K}_k{\cal H}) {\cal M} ({\bf 
x}^a_{k-1}) + { \bf K}_k{\bf y}^o_k.
\end{equation}
The repetition of these analysis and forecast steps is referred to as 
the {\it prediction-assimilation cycle}.

The effect of the observations can thus be interpreted as a forcing 
${ \bf K}({\bf y}^o-{\cal H}{\cal M} ({\bf x}^a))$, which acts on the 
free solution at the observation times $t_k$; note that observations 
are typically not available at every time step of the discretized set 
of nonlinear partial differential equations \cite{bengtsson81,GhilMal91}.
Equation (5) governs the sequential estimation problem, {\it i.e.} 
the evolving estimate of the state of the system; 
${\bf y}^o-{\cal H}({\cal M} ({\bf x}^a))$ here is the {\it innovation vector}.

We consider now a perturbed trajectory that undergoes the same 
forecast and assimilation steps, with the same observations, as the 
reference trajectory of (5). The equation describing the linear 
evolution of perturbations $\delta{\bf x}^{f,a}(t_k)$ of this 
prediction-assimilation cycle is:
\begin{equation}
\delta{\bf x}^{a}_{k} = ({\bf I} - {\bf K}_k {\bf H}_{k} ) {\bf 
M}_{k-1} \delta{\bf x}^{a}_{k-1},
\end{equation}
where ${\bf H}_{k}={\bf H}({\bf x}_{k}^f)$ is the Jacobian of $\cal 
H$ at time $t_{k}$ and ${\bf M}_{k-1}$ the linearized evolution 
operator associated with ${\cal M}$ between $t_{k-1}$ and $t_{k}$. 
The term $({\bf I} - {\bf K} {\bf H})$ appearing in (6) reflects the 
effect of the forcing induced by the assimilation.
This term modifies the stability properties of the perturbative 
dynamics relative to (5), {\it i.e.} its Lyapunov exponents, with 
respect to those of the free system (1).

For the updates to drive the solution of Eq. (5) towards the correct 
solution of Eq. (1), the forecast-assimilation cycle (5) must be 
stabler than the pure-forecast system (1). Hence the Lyapunov 
exponents of Eq. (5) must be algebraically smaller than those of Eq. 
(1), which usually leads to its unstable subspace
being lower-dimensional as well. Complete stabilization by the 
updating process, {\it i.e.} total absence of positive Lyapunov 
exponents, is sufficient for the uniqueness of the solution of Eq. 
(5), as well as necessary for the convergence of this solution to the 
true state of the system. Such complete stabilization will drive 
analysis errors to zero in the absence of observational and system 
noise, and to the lowest-possible values when noise or nonlinear 
effects are present.

There are two means at our disposal in order to achieve this stabilization 
of a prediction-assimilation cycle: the design of the observational 
network, corresponding to the operator ${\cal H}$, and that of the 
assimilation scheme, resulting in a certain gain matrix ${\bf K}$; it 
is the product ${\bf KH}$ in Eq. (6) that provides the stabilizing 
effect of the forcing by the data. A. Trevisan and associates 
\cite{TU, UT, CTU} have proposed an efficient way to achieve this 
stabilization and improve the performance of the data assimilation method, 
by monitoring the unstable modes that amplify along a trajectory of the 
prediction-assimilation system. In their AUS approach, the basis of 
the subspace to which the analysis update is confined is given by the 
unstable directions of the system.

The AUS gain matrix {\bf K}, which differs from zero only on the 
unstable subspace at an update point, is given by:
\begin{equation}
{\bf K} = {\bf E}{\bf \Gamma}({\bf HE})^T[({\bf HE})\Gamma({\bf 
HE})^T+{\bf R}]^{-1}.
\end{equation}
Here ${\bf E}$ is the unitary matrix
whose columns are the $m$ unstable directions, while ${\bf \Gamma}$ 
is a symmetric, positive-definite matrix representing the 
forecast-error covariance in the subspace spanned by the columns of 
${\bf E}$; the index $k$ is omitted in (7) for clarity. Since 
typically $m << n$, this feature of the method is clearly efficient 
in reducing the computational cost of the estimation process. A 
traditional, fixed network of observations can then be used to detect 
and reduce the forecast error projection along the unstable 
directions. An {\it adaptive observational network}, designed to 
measure primarily the unstable modes, will further enhance the 
efficiency of the assimilation.

The unstable directions of a dynamical system of type (1) can be 
estimated by the {\it breeding method} \cite{TothKalnay93, 
Patil-etal01}. In this procedure, the full nonlinear system is used 
to evolve small perturbations and, at fixed time intervals, their 
amplitude is scaled down to the initial value.
The extension of the breeding technique to Eq. (5), referred to as 
Breeding on the Data Assimilation System (BDAS) \cite{TU}, allows one 
to estimate the unstable directions of a prediction-assimilation 
system, subject to perturbations that obey Eq. (6).
In the AUS assimilation, the unstable directions are used in the 
definition of the matrix {\bf K}, cf. Eq. (7), and they can also be 
used to identify adaptive observations that are most beneficial for 
error reduction.

\section{Results}

\subsection{Theoretical results}

We first provide a theoretical result that, under simplified 
circumstances, gives the mathematically rigorous condition for the 
observational forcing to stabilize the prediction-assimilation cycle 
(5). This result helps to clarify the theoretical underpinnings of AUS.
Consider a chaotic flow, with a single positive Lyapunov 
exponent, and restrict the system's true evolution to an unstable fixed 
point, so that $\mathbf{M}$ in Eq. (6) is a constant matrix. 
The eigenvalues of this matrix are $\Lambda_i=e^{(\lambda_i\tau)}$,
where $\lambda_i$ are the Lyapunov exponents, 
and the eigenvectors of $\mathbf{M}$ are the Lyapunov vectors of the
flow, while $\tau=(t_k-t_{k-1})$ is the assimilation interval.
Alternatively, the result applies to the map associated with integer
multiples of the period along an unstable periodic orbit of such a flow.

Let the state of the system be
estimated using a single noisy observation at each analysis
time, assimilated by AUS along the single unstable direction $\mathbf{e}_{k}$,
so that: $\mathbf{K}_{k}=c_{k}\mathbf{e}_{k}$ and $c_{k}=\gamma 
^{2}\left(\mathbf{He}_{k}\right) \left[ \gamma 
^{2}\left(\mathbf{He}_{k}\right)^{2}+\sigma _{o}^{2} \right]^{-1}$,
$\gamma ^{2}$ and $\sigma _{o}^{2}$ being the forecast error variance
along $\mathbf{e}_{k}$ and the observation error variance, respectively.
\paragraph*{Theorem.} Let the constant matrix $\mathbf{M}$ have a single
eigenvalue $\Lambda >1$ corresponding to a positive Lyapunov exponent,
with $\mathbf{e}$
its associated eigenvector. Let $\mathbf{H}$ be a constant row vector, and the
Kalman gain be approximated by $\mathbf{K}_{k}=c\mathbf{e}_{k}$,
where $c$ is a constant scalar. The sequence ${\{\bf e}_k: k = 1, 2, 
...\}$ is defined by the recursion:
\begin{align*}
\Lambda _{k}\mathbf{e}_{k} & = \mathbf{Mf}_{k-1}, \tag{8a}\\ 
\Theta_{k}\mathbf{f}_{k} & = \left(\mathbf{I}-c\mathbf{e}_{k}\mathbf{H}\right)\mathbf{Mf}_{k-1}; \tag{8b} 
\end{align*}
here the initial $\mathbf{f}_{0}$ is an arbitrarily chosen unit
column vector, while $\Lambda _{k}$ and $\Theta _{k}$ are the normalization
factors associated with $\mathbf{e}_{k}$ and $\mathbf{f}_{k}$, respectively.
Then, a sufficient condition for the solution of (5) to be stable is:
$$c\mathbf{He}>1-{\Lambda}^{-1}. \eqno{(9)}$$
\paragraph*{Remark.} The amplitude $c$ of the correction, and the observed 
(scalar) component $\mathbf{He}$ of the unstable vector, must thus
be large enough to counteract the unstable growth.

\paragraph*{Proof.} It can be shown that both $\mathbf{e}_{k}$ and 
$\mathbf{f}_{k}$
converge to $\mathbf{e}$. The range of $\mathbf{K}$ is one-dimensional and
the eigenvector $\mathbf{e}$ of $\mathbf{M}$, with associated
eigenvalue $\Lambda $, is also an eigenvector of 
$\left(\mathbf{I}-\mathbf{KH}\right)\mathbf{M}$,
with associated eigenvalue $\Theta $ \[
\left(\mathbf{I}-c\mathbf{eH}\right)\mathbf{Me}=\left(1-c\mathbf{He}\right)\Lambda\mathbf{e} 
={\Theta}\mathbf{e}. \]
The stability condition is then obtained by setting $\Theta <1$, and 
this condition also guarantees uniqueness of the solution. $\quad\square$

To illustrate the essence of the theorem we give here a simple 
numerical example in the context of the three-variable Lorenz model 
\cite{Lorenz63}. When the canonical values of its parameters: $\sigma 
= 10$, $r = 28$ and $b = 8/3$ are chosen, the system behaves 
chaotically. In the example that follows, the "true" state of 
the system that we want to estimate is the phase-space origin $x = y 
= z = 0$, which is an unstable fixed point for the system. Two 
assimilation experiments are performed; each one evolves 
according to the observationally forced, discrete dynamical system 
(5). At each analysis time a single noisy observation, the 
$y$-variable, is assimilated.

The time step for integration has been set equal to $0.01$ time 
units, while the assimilation is performed every $10$ time steps, 
{\it i.e.}, $\tau=0.1$ time units; see also Miller et al. \cite{Milleretal94}. 
The matrix ${\bf M}$ represents the tangent linear operator 
evaluated at the origin and integrated for the time interval $\tau$.
 This matrix possesses exactly one eigenvalue larger than one, $\Lambda 
= 3.26$. The Kalman gain matrix, used to update the analysis, is ${\bf 
K}_k=c{\bf e}_k$, with ${\bf e}_k$ obtained by the recursive use of 
Eq. (8). The sequence ${\{\bf e}_k\}$ converges to 
the eigenvector ${\bf e}$ corresponding to the eigenvalue $\Lambda$, 
and the asymptotic value of ${\bf He}_k$ is ${\bf He}=0.91$. The 
stability condition (9) yields therefore $c > c_0=0.762$.

Figure 1 shows the root-mean-square (RMS) analysis error as a function 
of time (in units of assimilation interval, ${\it i.e.}$,  $\tau=10$ time steps) for 
the two experiments, with $c = 0.76$ and $c = 0.77$, respectively. The 
observational RMS error is $10^{-3}$. Clearly the RMS grows exponentially 
in the former case and decays exponentially in the latter.


\subsection{Numerical results}

We now illustrate the stabilizing effect of observational forcing for 
different assimilation schemes and show that their long-term 
performance is related to the induced degree of stabilization. The 
full Lyapunov spectrum is used to compare the properties of a given free 
system (1) with those of the corresponding prediction-assimilation system (5). 
Observing system simulation experiments \cite{bengtsson81,GhilMal91,Ghil97} 
are performed with numerical models of increasing complexity.

The first chaotic model \cite{LorenzEm98} has 40 scalar 
variables that represent the values of a meteorological field at 
equally spaced sites along a latitude circle. It can be derived from an
``anti-Burgers" equation in the same way that the Fermi-Pasta-Ulam model
was subsequently shown to be derived from the Korteweg-deVries equation \cite{FPU50}.
The second is an atmospheric model that represents mid-latitude, 
large-scale flows, and is based on the quasi-geostrophic equations \cite{kalnay03}
in a periodic channel. The discretized model has 15 000 scalar variables
and its details can be found in \cite{RotBao96}, 
while  the experimental setup for assimilating data is described in 
\cite{CTU}.

Figure 2 shows the spectrum of the 40 Lyapunov exponents of the 
first model \cite{LorenzEm98}, using 200 years of simulation. 
The free system possesses 13 positive exponents ($\lambda^{+}$), of 
which the leading one ($\lambda_{max}$ = 0.336 day$^{-1}$) 
corresponds to a doubling time of 2.06 days; its Kaplan-Yorke 
dimension is approximately 27.05.
At each assimilation time, the analysis is performed by AUS with a 
single BDAS mode ${\bf e}^{BDAS}_k$ in the specification of the gain 
matrix, Eq. (7). In this single-observation situation, the matrices 
${\bf E}$ and ${\bf \Gamma}$ reduce to a column vector and to a 
scalar, respectively; the latter is estimated statistically using the 
innovations --- which are scalars, in the present case --- following the approach of \cite{CTU}.

A single observation, adaptively located where the current BDAS mode 
attains its maximum value, is sufficient to stabilize the system, so that 
$\lambda_{max}$ = $-0.283$ day$^{-1}$, and to reduce the RMS analysis 
error to 1.4$\%$ of the system's natural variability, even when the assimilation 
interval $\tau=t_{k+1}-t_k$ is as long as 3 hr. Smaller assimilation intervals 
of 2 hr and 1 hr lead to further stabilize the system 
($\lambda_{max}$ = $-0.437$ day$^{-1}$ and $-0.809$ day$^{-1}$, 
respectively), and thus to reduce the analysis error even more: the RMS 
error, normalized by the natural variability, becomes 0.011 and 
0.009, respectively.


Figure 3 shows the first 100 Lyapunov exponents of the 
quasi-geostrophic model \cite{RotBao96}, using one year of simulated 
time. The free system possesses 24 positive exponents 
($\lambda^{+}$), of which the leading one ($\lambda_{max}$ = 0.310 
day$^{-1}$) corresponds to a doubling time of 2.2 days, and its 
Kaplan-Yorke dimension is approximately 65.2. The three assimilation 
experiments all use a fixed network of noisy observations that cover 
just under one-third of the domain (20 out of 64 meridional lines of 
grid points); in two of the experiments, an additional observation is 
adaptively located at a single grid point in the otherwise unobserved 
portion of the domain. Its location coincides with the maximum of a 
single BDAS mode, as  in the first model \cite{LorenzEm98}.
The model time step is $\Delta t = 30$ min, while the assimilation 
interval $\tau$ is 6 hr.

In all three experiments, the fixed observations are assimilated by a 
least-square fit, according to the three-dimensional variational 
(3DVar) algorithm \cite{talagrand97,kalnay03} in wide operational use, while 
the adaptive observations are assimilated either by 3DVar 
(3DVar-BDAS) or in the unstable subspace (AUS-BDAS)
by using the current BDAS mode in Eq. (7). When fixed observations 
only are assimilated (3DVar), the number of positive exponents is 
reduced to three, with the leading exponent ($\lambda_{max}$ = 0.088 
day$^{-1}$) corresponding to a doubling time of 7.9 days, while the 
Kaplan-Yorke dimension is reduced to 6.9 and the normalized RMS error 
to 0.321. Adding a single adaptive observation assimilated by 3DVar 
(3DVar-BDAS) stabilizes the prediction-assimilation cycle further, 
the only positive exponent being slightly greater than zero 
($\lambda_{max}$ = 0.002 day$^{-1}$, Kaplan-Yorke dimension 1.1), and 
reduces the normalized RMS analysis error to around 0.16. Finally, 
when the adaptive observation is assimilated in the unstable 
subspace, the system is completely stabilized and the RMS analysis 
error drops to only 0.058.


\section{Conclusion}
To estimate the efficiency of prediction-assimilation systems and observational 
networks, one often estimates the error of a short-range forecast at 
points where fairly accurate observations are available; the obvious 
drawback of this approach is that errors tend to be smaller in 
systematically observed regions \cite{bengtsson81, GhilMal91, 
Ghil-etal81}. The nonlinear stability analysis introduced here allows 
one to address these issues in a more rigorous way.  The stability of the 
prediction-assimilation system guarantees the uniqueness 
of its solution and is required for the convergence of this solution to the 
true flow evolution; in turn, the degree of stabilization introduced 
by the data assimilation may be measured precisely by estimating the 
full Lyapunov spectrum of the forced system.

Assimilation in the unstable subspace and the use of breeding on the 
prediction-assimilation system to estimate this subspace were applied here 
to a 40-variable \cite{LorenzEm98} and to a 15 000-variable model 
\cite{RotBao96} simulating the mid-latitude atmospheric circulation; 
in both models, the proposed methods led to complete stabilization of 
the sequential estimation process.
The tools of dynamical systems theory can thus help design optimized 
assimilation algorithms (the specification of {\bf K}) and 
observational networks (specifying ${\cal H}$) for high-dimensional and highly 
nonlinear systems. When, as usual in practical applications, only a 
limited number of measurements can be made, asymptotic error 
reduction may still be achieved through adaptive deployment of 
observations  and a sophisticated assimilation scheme, designed to 
control the flow's instabilities.

Data assimilation applications are possible in all situations where a 
dynamical constraint is important and only a limited amount of noisy 
observations can be taken. Such situations include robotics, flow in 
porous media, plasma physics, as well as solids subject to thermal 
and mechanical stresses or to shocks \cite{kao-etal04, kao-etal06}. 
The insights into the nonlinear dynamics of the 
prediction-assimilation cycle provided here should help design data 
assimilation methods for a large class of laboratory experiments 
\cite{galmicheetal03}, as well as for natural or industrial systems.

\section*{Acknowledgments}
We thank Micka{\"e}l Chekroun for his insightful comments and 
suggestions. This work was supported by the Belgian Federal Science 
Policy Program under contract MO/34/017 (AC), by Cooperazione 
Italia-USA 2006-2008 su Scienza e Tecnologia dei Cambiamenti 
Climatici (AT) and by NASA's Modeling, Analysis and Prediction (MAP) 
Program project 1281080 (MG).

\newpage
\begin{figure}
\includegraphics[width=8cm, height=6.6cm]{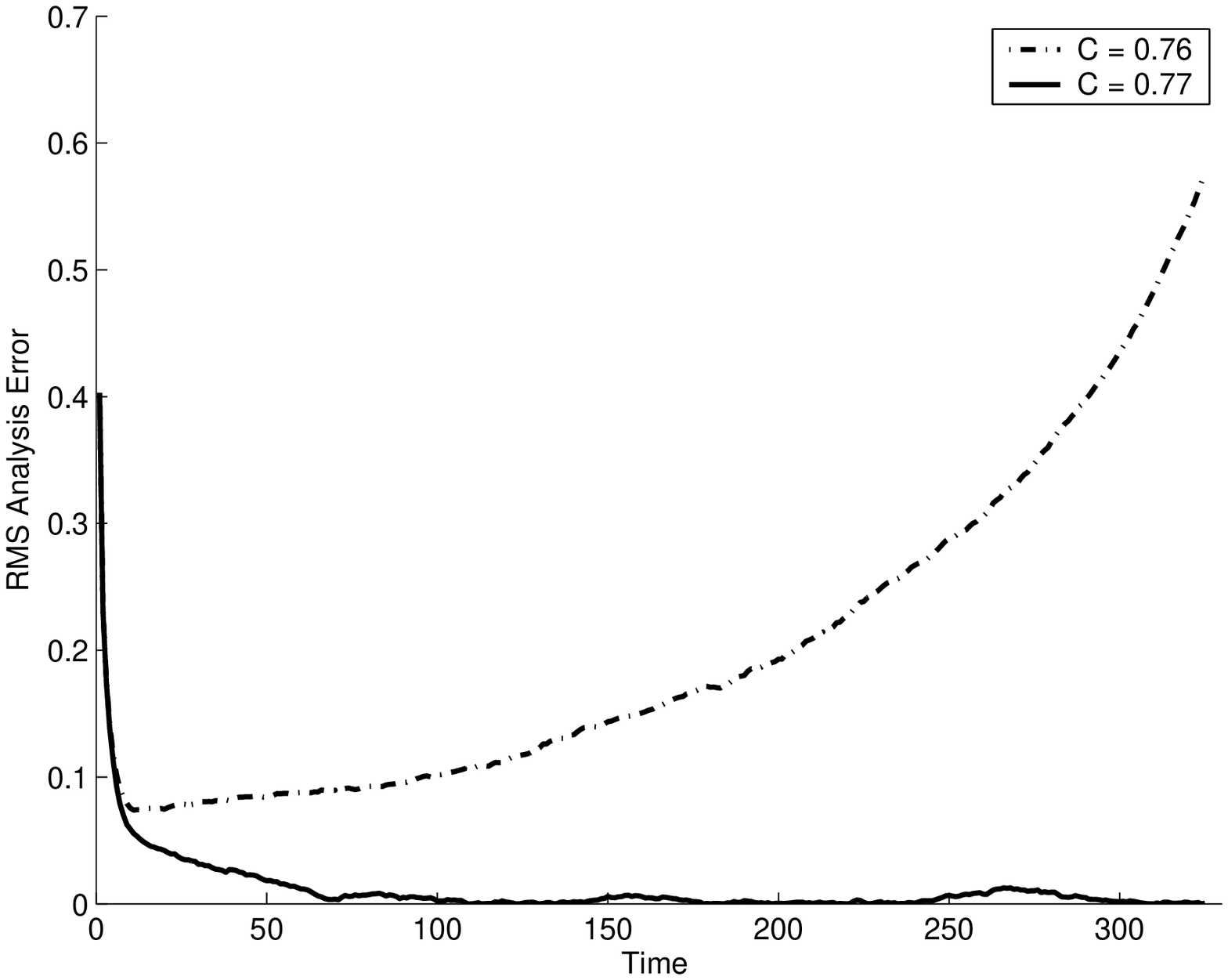}
\caption{ RMS analysis error as a function of time for two 
experiments with $c = 0.76$ and $c = 0.77$, below and above the 
stability threshold $c_0=0.762$; the time is given in multiples of 
the updating interval $\tau = 0.1$.}
\end{figure}

\begin{figure}
\includegraphics[width=8cm, height=6.6cm]{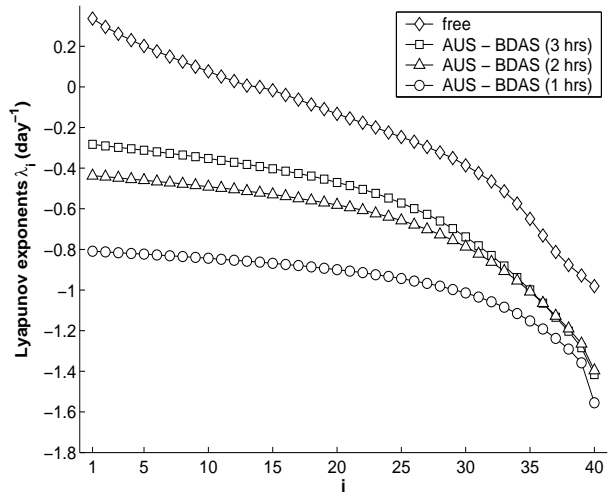}
\caption{\label{FIG1} Effect of the observing system set-up on 
stability of the prediction-assimilation cycle for the Lorenz 
40-variable model \cite{LorenzEm98}. Spectrum of the Lyapunov 
exponents for the free system (1) (diamonds) and for the AUS-BDAS 
forced system (5), given different assimilation intervals 
$\tau=t_{k+1}-t_k$: 3 hr (squares), 2 hr (triangles) and 1 hr 
(circles).}
\end{figure}

\begin{figure}
\includegraphics[width=8cm, height=6.6cm]{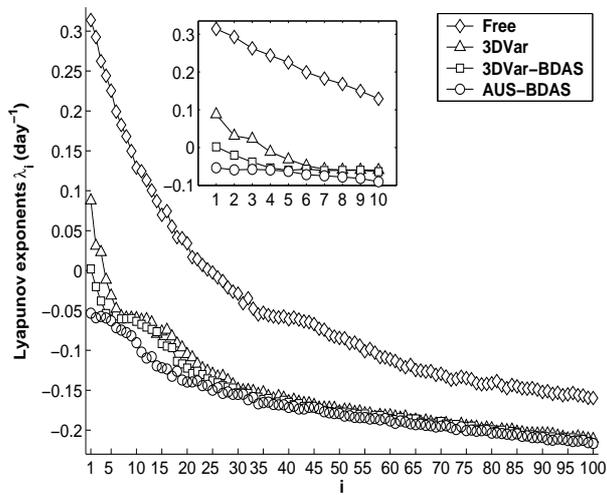}
\caption{\label{FIG2} Effect of the assimilation method on stability 
for the quasi-geostrophic model \cite{RotBao96}. Spectrum of the 
first 100 (the first 10 in the inset) Lyapunov exponents for the free 
system (1) (diamonds) and for the observationally forced systems (5): 
3DVar (squares), 3DVar-BDAS (triangles) and AUS-BDAS (circles).}
\end{figure}

\end{document}